\documentstyle[12pt,aps,prl,floats,eqsecnum,graphicx,euscript,amsmath]{revtex}

\newcommand{\sign}{\mathop{\rm sign}\nolimits}

\def\lapprox{\,\raise0.4ex\hbox{$<$}\kern-0.8em\lower0.7ex\hbox{$\sim$}\,}
\def\gapprox{\,\raise0.4ex\hbox{$>$}\kern-0.8em\lower0.7ex\hbox{$\sim$}\,}

\begin{document}

\bibliographystyle{prsty}
\centerline{\large\bf Spin Relaxation in the Quantized Hall Regime}
\centerline{\large\bf in the Presence of Disorder}
\vskip 3mm
{
\centerline{S. Dickmann}
}

\centerline{\it Institute for Solid State Physics of RAS, Chernogolovka, 142432 Moscow District, Russia}

\centerline{\it and Weizmann Institute of Science, 76100 Rehovot, Israel}
\vskip 5mm

\begin{abstract}
We study the spin relaxation (SR) of a two-dimensional electron gas (2DEG) in the
quantized Hall regime and discuss the role of spatial inhomogeneity effects
on the relaxation. The results are obtained for small filling factors ($\nu\ll
1$) or when the filling factor is close to an integer. In either case SR times
are essentially determined by a smooth random potential.
For small $\nu$ we predict a ``magneto-confinement" resonance manifested in
the enhancement of the SR rate when the Zeeman energy is close to the
spacing of confinement sublevels in
the low-energy wing of the disorder-broadened Landau level. In the resonant
region the $B$-dependence of the SR 
time has a peculiar non-monotonic shape. If
$\nu\simeq 2n\!+\!1$, the SR is going non-exponentially. 
Under typical conditions the calculated SR times range from $10^{-8}$ to $10^{-6}\,$s.
\end{abstract}
\vskip 5mm

{\bf 1.} For the relaxation of electron spins to occur, two conditions
have to be fulfilled:
the first one is the presence of an interaction mixing different spin states
in the system studied; the second is the availability of a mechanism which makes
the relaxation process irreversible. Both conditions can be realized in a
rich variety of ways,
and even in the case
of two-dimensional (2D) electrons one finds a wide scatter
of experimental data \cite{do88} and theoretical results
\cite{by84,dy86,fr91,ba92,di96,di99,kh00,kh01,al01} devoted to SR
problems.
Besides, in the magnetic field the SR process is actually the relaxation
of the Zeeman energy
$|g\mu_B B\Delta S_z|$ (${\vec B}\parallel {\hat z}$, $\Delta S_z=S_z-S_0$
is the spin
deviation of the $S_z$ component from the equilibrium value $S_0$).

Theoretically, the relaxation problem of a flipped spin in a
semiconductor heterostructure in high perpendicular magnetic field 
seems to be first
formulated by D. Frenkel \cite{fr91},  who considered
the relativistic part
of the phonon field acting directly on the electron spin as
the interaction mixing the electron spin states.
One could
estimate that in the case of another mechanism, exactly due to
the spin-orbit (SO) coupling reformulated for the 2D case
\cite{by84,dy86}, the relevant spin-flip transition matrix element is
at least by an order of magnitude greater. However,
the work of Ref. \cite{fr91} caused some misunderstandings even almost a
decade after its publication. (See the comment \cite{kh01} and
references therein).

Any properties of a 2DEG in the quantum Hall regime crucially
depend on the filling factor  $\nu=N/N_{\phi}$, where $N$ and
$N_{\phi}=L^2/2\pi l_B^2$ are the numbers of electrons and
magnetic flux quanta, respectively ($L\times L$
is the 2DEG area, $l_B$ is the magnetic length).  In this Letter
we consider the SR problem in two formulations. 

First, we
solve {\it the problem of one-electron spin-flip}
in the presence of a random potential, and in so doing we study
the SR in a lateral quantum dot in high magnetic field, where
the effects of interplay between the localization and the Zeeman
coupling are essential. We emphasize that we deal with a weak
confinement but in the presence of strong $B$. Therefore 
well smaller energies are relevant ($\sim\!1\,$K) than,
e.g., in the case of usual interplay of Fock-Darwin states \cite{fo28}. (The
latter is based on competition of confinement and
cyclotron energies which are $\gapprox 10\,$K). The most striking
manifestation of the {\it ``magneto-confinement"} effect occurs when
the Zeeman energy is close to the spacing of lowest-energy levels
in a quantum dot or in a potential minimum. This results in  
{\it strong enhancement of the SR rate}.
The studied one-electron problem is actual
for small filling factors, $\nu\ll 1$, but in strong magnetic
fields it may also be extended in terms of the Hartree-Fock
approach to a vicinity of an even filling factor, when $\nu\simeq
2n$.

Second, we report on the results for the SR in {\it a
strongly correlated} \mbox{2DEG} when it presents {\it a quantum
Hall ferromagnet} (QHF), i.e. when the filling factor $\nu$  is
close to an odd integer ($\nu\simeq 2n+1$).  In this
case we study a new mechanism of relaxation of the total 2DEG
spin, where the cause of irreversibility is neither
electron-phonon interaction \cite{di96} nor
inter-spin-wave scattering \cite{di99} but a disorder.  We
will see that in real experimental regimes this disorder
relaxation channel has significantly to prevail over the phonon one.  

In both SR problems the temperature is assumed to be equal to
zero which actually means that it is lower than the
Zeeman energy.

{\bf 2.} We consider a smooth random potential (SRP) as the
disorder.  The total single-electron Hamiltonian is thereby as follows:
$
  {\EuScript H}= \hbar^2\hat{{\bf q}}^2/2m_e^* - \epsilon_Z{\hat \sigma}_z/2 
  + u({\bf r})+H_{SO}  
  + U_{\mbox{\scriptsize e-ph}},    
$
where $\epsilon_Z=|g|\mu_BB$, $\hat{{\bf q}}=-i{\bf \nabla}+
e{\bf A}/c$ and ${\bf r}=(x,y)$
are 2D vectors, $u({\bf r})$ is the SRP field, the $H_{SO}$ and 
$U_{\mbox{\scriptsize e-ph}}$ terms respond to the SO and electron-phonon
interactions (see below).
If the SRP is assumed to be Gaussian, then it is
defined by the correlator $K({\bf r})=\langle u({\bf
r})u(0)\rangle$. We choose also $\langle u({\bf r})\rangle=0$
which means that the SRP energy is measured from the center of the Landau
level. In terms of the correlation length $\Lambda$ and
Landau level width $\Delta$, the correlator is
$$
  K({\bf r})=\Delta^2\exp{(-r^2/\Lambda^2)}\,.   \eqno (1)
$$
In the realistic case $\Delta\sim 10\,$K, $\Lambda\sim
30-50\,$nm, therefore $\Delta \gapprox \hbar^2/m_e^*\Lambda^2$. 
We study the case $\Delta\ll \hbar\omega_c$ ($\omega_c$ is the
cyclotron frequency), and $\Lambda\gg l_B$.  
In the SRP field the electron drifts quasiclassically along
an equipotential line. However before the spin-flip it relaxes
to a SRP minimum.  Estimates for this relaxation time (due to phonon
emission without any spin-flip) yield the values not exceeding $1\,$ns.

For simplicity, we model the SRP in the vicinity of a minimum by a parabolic
confinement potential $u=m_e^*\omega^2r^2/2$, and to describe
the electron states
we use the symmetric gauge basis (${\bf A}={\bf r}\!\times\!{\bf B}/2$):
$$
  |n,m,\sigma\rangle=
   \sqrt{\frac{n!}{(n+m)!}}
  \frac{e^{-im\varphi-r^2/4a^2}}{a\sqrt{2\pi}}
  \left(\frac{ir}{\sqrt{2}a}\right)^{m}L^m_n(r^2/2a^2)|\sigma\rangle\,.
  \eqno (2)
$$
($L^m_n$ is a Laguerre polynomial; note also that only the
states with $n\!+\!m\ge 0$ are considered in the following.) For the
length $a$ it should be substituted $a\!=\!(\hbar/2m_e^*\Omega)^{1/2}$,
where  $\Omega=\sqrt{\omega^2\!+\!\omega_c^2/4}$. The system thus 
becomes equivalent to a lateral quantum dot \cite{kh00,al01},
and we deal with the Fock-Darwin states \cite{fo28} with
energies $E_{n,m}=\hbar(2n\!+\!m\!+\!1)\Omega-
\hbar\omega_cm/2$.
The appropriate quantity is also the level spacing
$\delta=\hbar(\Omega\!-\!\omega_c/2)\approx
\hbar\omega^2/\omega_c$, concerning the $m\ge 0$ states which
belong to the same number $n$. We calculate the 
{\it total rate} of the
transition of an electron initially occupying
the upper spin sublevel to {\it any
final state} of the lower spin sublevel. At first sight
we should
consider the spin-flipped state $|0,0,\downarrow\rangle$ as
the initial one. However, a correction of the 
states due to the SO coupling has to be taken into account.
We use the SO Hamiltonian specified for the (001) GaAs plane: 
  $H_{SO}=\alpha\left(\hat{{\bf q}}\times\hat{\mbox{\boldmath $\sigma$}}
  \right)_{\!z}+
  \beta\left(q_y{\hat \sigma}_y\!-\!q_x{\hat \sigma}_x\right)$.
This expression is a combination of the Rashba term \cite{by84} (with the
coefficient $\alpha$) and the crystalline anisotropy term
\cite{dy86,ba92,di96,di99,kh00,kh01,al01}
($\hat{\sigma}_{x,y,z}$ are the Pauli matrices). Assuming that
$\alpha$ and $\beta$ are small ($\alpha, \beta\ll \hbar\omega_c
l_B$) we find after perturbative treatment the spin-orbitally 
corrected states before and after the spin flip:
$$
  |i\rangle=C_{1,1}|0,0,\downarrow\rangle+C_{2,1}|0,1,\uparrow\rangle+\frac{\beta}{\hbar\Omega
  \sqrt{2}a}|1,-1,\uparrow\rangle\quad\mbox{and} \eqno (3)
$$
$$
  |f_m\rangle=C_{1,m}|0,m,\uparrow\rangle+C_{2,m}|0,m\!-\!1,\downarrow\rangle-
  \frac{i\alpha}{\hbar\Omega
  \sqrt{2}a}|1,m\!-\!1,\downarrow\rangle+\frac{\beta\delta\sqrt{m\!+\!1}}{\hbar\Omega
  \sqrt{2}a(\delta+\epsilon_Z)}|1,m\!+\!1,\downarrow\rangle, \eqno (4)
$$
where $0\le m<\epsilon_Z/\delta$ (the Zeeman energy $\epsilon_Z$ has been neglected
as compared to $\hbar\omega_c$). The coefficients $C_{i,m}$ are
defined as follows: let
$T=\alpha\delta\sqrt{2}/a\hbar\Omega(\epsilon_Z-\delta)$ and
$P_m=2\sqrt{1\!+\!T^2m}$, then $C_{1,m}=\sqrt{1/2\!+\!1/P_m}$,
and $C_{2,m}=i\sign{(\delta-\epsilon_Z)}\sqrt{1/2\!-\!1/P_m}$.
Here the resonance mixing of the ``spin-up'' and ``spin-down''
states (if $\epsilon_Z\simeq \delta$) has been properly taken
into account. Note that the behaviour of the states (3)-(4) 
in the vicinity of the
resonance (namely in the interval $\Delta
B/B\lesssim\alpha\sqrt{m_e^*/\hbar^3\omega_c}\lesssim 0.1$) is
governed only by the Rashba SO mechanism.

In the resonance region $C_{1,1}\sim |C_{2,1}|$, and $|i\rangle$
is thereby a well hybridized spin state. The
$|i\rangle\!\to\!|f_0\rangle$ transition then is not due to the
SO coupling, and this should lead to the SR enhancement.
The final state $|f_0\rangle$ is the almost ``pure'' spin-state.
In fact we may always set
$C_{1,m}=1$ and $C_{2,m}=0$ in the expression for $|f_m\rangle$.
Indeed, though the spin hybridization of the $|f_1\rangle$ state is
significant, it plays negligible role in the SR process because
of vanishing of the relevant phonon momentum
$(\epsilon_Z\!-\delta)/c_s$. Then we find the matrix element $\langle
f_m|U_{\mbox{\scriptsize e-ph}}|i\rangle$ and, using the Fermi
golden rule, obtain the SR rate within a certain SRP minimum,
$1/\tau_{\omega}=\frac{2\pi}{\hbar}\sum_{m,{\bf k}}\left|\left\langle f_m\left
  |U_{\mbox{\scriptsize e-ph}}\right|i\right\rangle\right|^2
  \delta(\hbar
  c_s k-\epsilon_Z).  
$
Here $U_{\scriptsize{\mbox{e-ph}}}({\bf r})=
(\hbar/V)^{1/2}\sum_s
{\tilde U}_s({\bf q},k_z)e^{i{\bf q}{\bf r}}$,
where the index $s$ labels phonon polarization, $V$ is the sample
volume, and ${\tilde U}_s$ is the renormalized (in the 2D layer)
vertex which includes the deformation and piezoelectric fields
created by the phonon \cite{di96,gale87}. The summation over $s$ involves averaging over
directions of the polarization unit vector for both
components of the electron-phonon interaction, and this may be
reduced to $\left|\sum_s U_s\right|^2={\pi\hbar
c_s k}/p_0^3\tau_A({\bf k})$, where ${\tau_A}^{-1}=\tau_D^{-1} +
5\tau_p^{-1}{p_0^2}(q^2k_z^2+q_x^2q_y^2)/k^6$ (see Ref. \cite{di96}).  The nominal
times for the deformation and piezoelectric interactions in GaAs
are $\tau_D\approx 0.8\,$ps, and $\tau_P\approx 35\,$ps
\cite{di96,gale87}. The nominal momentum is $p_0=2.52\cdot
10^6/$cm \cite{gale87}. (We also refer to Refs.
\cite{di96,gale87} for details concerning the meaning of these
quantities and their expressions in terms of the GaAs
material parameters.) Finally, in the general expression for
$1/\tau_{\omega}$ we perform  the
summation and after a routine treatment arrive at the result
$$
  \frac{1}{\tau_{\omega}}=\frac{1}{8ap_0\tau_p}\int_0^1\frac{d\xi}
  {\sqrt{1\!-\!\xi}}
  \sum_{0\le m<{\epsilon_Z}/{\delta}}\frac{b_m}{m!}\left[{\cal A}_m^2(\xi)+
  {\cal B}_m^2(\xi)\right](S_m+5\xi-35\xi^2/8)e^{-\xi b_m^2}
  \left(\frac{\xi b_m^2}{2}\right)^{\!\!m}\!\!,       \eqno (5)
$$
$$
  \mbox{where}\;\,
  b_m\!=\!a\frac{\epsilon_Z\!-\!m\delta}{\hbar c_s},\;\; S_m=
  \frac{b_m^2\tau_p}{(ap_0)^2\tau_d},
  \;\;
  {\cal A}_m\!=\!\left(\frac{2m}{b_m\sqrt{\xi}}\!-\!\sqrt{\xi}b_m\right)|C_{2,1}|\!+\!
  \frac{\alpha b_m}
  {\hbar\Omega a}\sqrt{\frac{\xi}{2}}\sign{(\delta\!-\!\epsilon_Z)},
$$
and ${\displaystyle {\cal B}_m\!=\!
  \frac{\beta b_m}{\hbar\Omega a}\sqrt{\frac{\xi}{2}}
  \left(1\!-\!C_{1,1}\frac{\delta}{\delta\!+\!\epsilon_Z}\right)}$. The
  estimate for $\alpha$ and $\beta$
depends on the effective layer width
\cite{by84,dy86}. The rate $1/\tau_{\omega}$ as a function of $B$ at
$\omega=5\,$K is shown in the inset of Fig.1 for realistic parameters
indicated in the capture. 

At $\omega=0$ (i.e. in the ``clean" limit), the summation in the formula (5) 
is carried out over all numbers $m\!=\!0,1,...\infty$, and the expression (5) 
is reduced to 
$$
  \frac{1}{\tau_0}=
{\cal V}\int_0^1\frac{\xi\,d\xi}{\sqrt{1-\xi}}e^{-b^2\xi/2}(S+5\xi-35\xi^2/8)\,,
\eqno (6)
$$
where
${\displaystyle {\cal V}=\frac{{\epsilon_Z}^3(\alpha^2+\beta^2)}{4p_0\hbar^5\omega_c^2
  c_s^3\tau_p}\propto B},{}\;$
${\displaystyle b=\frac{\epsilon_Zl_B}{\hbar c_s} \propto \sqrt{B}},{}\;$
${\displaystyle S=\frac{\tau_p\epsilon_Z^2}{\tau_D(\hbar c_sp_0)^2}\propto B^2}$. 
The bold curves in the inset and in the main picture of Fig. 1 show
the corresponding $B$-dependencies. 

Note that if $\epsilon_Z=0$, then at any ${\bf r}_0$ the projection 
${\cal P}({\bf r}_0)=
\langle f_m|\delta({\bf r}\!-\! {\bf r}_0)
|i\rangle$ vanishes when 
calculated at a finite $\omega$ in the
leading order in the SO constants. [It does not occur for the
``clean" states, i.e. if in Eqs. (3) and (4) we pass to the $\omega\!\to\! 0$ 
limit before equating  $\epsilon_Z$ to zero.] 
Such a vanishing is a manifestation of the
general feature \cite{kh00,al01}: at zero Zeeman energy the
effects of the SO coupling in the leading order and of the
orbital magnetic field are similar. In particular, in a quantum
dot (where $\delta > \epsilon_Z$!), the first order SO
approximation in the $\epsilon_Z\to 0$ limit results only in a
small rotation of eigen states in the spin space.
With decreasing $B$ we get ${\cal P}\propto B^{3/2}\,$
(if $\epsilon_Z\ll\delta\ll\omega_c$)
and  obtain a sharper fall of the relaxation rate as compared to
the ``clean'' case (6).

In the presence of the SRP we have to carry out averaging
$1/\tau_{\mbox{\scriptsize eff}}=\int_0^\infty d\omega
F(\omega)/\tau_{\omega}$, where the distribution function
$F(\omega)$ is the probability for the confinement
frequency to take a certain value $\omega$. One may prove that
in the case of a Gaussian potential $u({\bf r})$ it should be
chosen in the form
$F(\omega)=\frac{2\omega}{\sqrt{\pi}\omega_0}\exp{(-\omega^4/4\omega_0^4)}$.
[The value $\omega^2$ is proportional to the curvature $\nabla^2u$,
and a routine analysis yields
$\omega_0=6^{1/4}(\Delta/m_e^*)^{1/2}/\Lambda$.]
Calculating $1/\tau_{\mbox{\scriptsize eff}}$ with this
function, we obtain the final result (see Fig.1).  As it has to
be, in comparison with $\tau_{\omega}$ the resonant behaviour of
$1/\tau_{\mbox{\scriptsize eff}}$ is smoothed, but at actual
values of $\omega_0$ it
results in non-monotonic $B$-dependence of $\tau_{\mbox{\scriptsize
eff}}$. Beyond the resonance region the behaviour is as
follows:  (i) at small magnetic fields (when $\epsilon_Z\ll
\hbar\omega_0^2/\omega_c \ll \omega_c $)
only one final state $|f_0\rangle$ participates in the SR, and
we find that $\tau_{\mbox{\scriptsize eff}} \propto B^{-5}$;
(ii) at high fields (when $\epsilon_Z\gg \delta_0$) there is a
large but finite number of possible states $|f_m\rangle$ into
which the confined spin-flipped electrons could relax, and the SR
time is always longer than $\tau_0$ but approaches this with
increasing magnetic field. Note that exactly in this high-field
regime the one-electron model becomes relevant for fillings
$\nu\simeq 2n$. Then the total 2DEG spin is determined only by a
small amount $|\nu\!-\!2n|N_{\phi}$ of effectively ``free''
electrons/holes belonging to the $(n\!+\!1)$-st/$n$-th Landau
level.

{\bf 3.} So, the problem has been solved when the total 2DEG spin is well
smaller than $N_{\phi}$. Now we study the opposite case: in the
ground QHF state the spin numbers attain maximum,
$S\!=\!S_z\!=\!N_{\phi}/2$.  This case is also remarkable, since
to the first order in the ratio $r_c=(e^2/\varepsilon
l_B)/\hbar\omega_c$ the low-lying excitation are again exactly
known: these are 2D spin waves or spin excitons (SEs).  The most
adequate description of the SE states is realized by spin-exciton
creation $Q_{\bf q}^{\dagger}=\frac{1}{\sqrt{N_{\Phi}}} \sum_p
e^{-ipq_xl_B^2} a^{\dagger}_{\downarrow\,p\!+\!\frac{q_y}{2}}
a_{\uparrow\,p\!-\!\frac{q_y}{2}}$ and annihilation $Q_{\bf
q}=\left(Q_{\bf q}^{\dagger}\right)^{\dagger}$
operators \cite{dz83}. In this definition $a_{\sigma p}$ stands for
the Fermi annihilation operator corresponding to the
$|n,p,\sigma\rangle=L^{-1/2}e^{ipy}\varphi_n(x\!+\!pl_B^2)$ state in the
Landau gauge ($\varphi_n$ is the $n$th harmonic oscilatory function). So the
one-exciton state is $Q_{\bf q}^{\dagger}|0\rangle$, where
$|0\rangle$ stands for the ground state. At small 2D momentum (${
q}l_B\ll 1$) the one-exciton state has the energy ${\cal
E}_{q}\!=\!\epsilon_Z\!+\!(q\l_B)^2/2M_n$ (now we need only this
small momenta approximation, see also general expressions for the
2D magneto-excitons at integer filling factors in Refs.
\cite{by81}). $M_n$ is the SE mass at $\nu=2n+1$, namely in
the $r_c\!\to\!\infty$ limit: $1/M_0\!=\!(e^2/\varepsilon
l_B)\sqrt{\pi/8}$, $1/M_1\!=\!7/4M_0$,...  Note that the sum 
$S_-=\sum_i\sigma_-^{(i)}$ lowering the spin number $S_z$ by
1 ($i$ labels the electrons), when considered to be projected onto
the $n$th Landau level, is simply proportional to  the ``zero''
exciton creation operator $Q_0^{\dagger}$. When the
SO coupling is ignored, any state
$\left(S_-\right)^N\!\!|0\rangle$ is
the eigenstate independently of the $r_c$ magnitude and of the presence
of a disorder.

The SO interaction $H_{SO}$ and the SRP field $u({\bf r})$
may be accounted perturbatively as usual. Meanwhile, now the unperturbed
part of the Hamiltonian involves the Coulomb interaction.
For present purposes the approximation in the context of
the projection onto a single Landau level is quite sufficient
(c.f. Ref. \cite{di96}). We note also that the
QHF state $|0\rangle$ is resistant to the SRP disorder. Such a stability
is determined by the exchange energy ($\sim e^2/\kappa l_B$) which 
is much larger than the amplitude $\Delta$.   

There are two fundamentally different alternatives to provide the
initial perturbation of spins. The first one is the perturbation of the
spin system as a whole when the $S$ number is not changed: $\Delta
S\!=\!0$, but $\Delta S_z\!\ne\! 0$. This is a Goldstone mode
which presents a {\it quantum precession} of the vector ${\bf S}$ around
the ${\bf B}$ direction. In terms of the SE representation $|\Delta
S_z|\!=\!N_{\phi}/2\!-\!S_z$ is the number of ``zero'' SEs excited
in a 2DEG. Let $|\Delta S_z|\!=\!N$; the corresponding state is
$\left(S_-\right)^N\!\!\left|0\right\rangle\!\propto\!
\left(Q_0^{\dagger}\right)^{N}\!\!\left|0\right\rangle$, and it
has the energy $N\epsilon_Z$. [Note that ``zero'' SEs do not
interact among themselves; besides, the stability of the 
$\left(S_-\right)^N\!\!|0\rangle$ state with respect to the disorder 
is identical to that for the $|0\rangle$ state because of $u({\bf r})$ 
and $S_-$ commuting.] 
The second case of the perturbation is
the $\Delta S\!=\!\Delta S_z$ type of the deviation. This does not
change the symmetry and involves the excitation of ``nonzero''
SEs, where each SE changes the spin numbers by 1:
$S\!\to\!S\!-\!1$, $S_z\!\to\!S_z\!-\!1$. In contrast to ``zero''
SEs, the ``nonzero'' ones interact. This interaction 
\cite{di99} or/and the direct exciton-phonon coupling \cite{di96}
govern the ``nonzero" SE annihilation which should go faster than for
``nonzero" SE annihilation process.

The SRP inhomogeneity does not
essentially affect the SE energy and the ``nonzero" annihilation. 
Indeed, the
exciton is neutral, and the interaction with the SRP incorporates
the energy $U_{\mbox{{\scriptsize x-}{\tiny SRP}}}\sim 
ql_B^2\Delta/\Lambda$ (the ``nonzero" SE possesses the dipole momentum
$el_B^2[{\bf q}\!\times\!{\hat z}]$, see Ref. \cite{by81}). The latter may be 
negative but is always
smaller than the ``nonzero'' SE energy ${\cal E}_q$. If the SE would
annihilate, the energy conservation condition ${\cal E}_{q}+
U_{\mbox{{\scriptsize x-}{\tiny SRP}}}=0$ is not satisfied except
of negligibly rare points, where the gradient $\nabla u$ is
accidentally large. Therefore, the SRP leads only to small corrections
(of the order of $U_{\mbox{{\scriptsize x-}{\tiny SRP}}}/{\cal
E}_{q}$) to the other mechanisms of the ``nonzero'' SE
relaxation \cite{di96,di99}.

A distinctly different process contributes to the SR in the case
of the first type of the spin perturbation. This is an effective
interaction of ``zero'' SEs among themselves arising due to
the SO coupling \cite{di96}. Such an interaction does not preserve
the total number of excitons $N$: at elementary event two
``zero'' excitons merge into one ``nonzero'' (the spin momenta
change following the rule $S_z\!\to\!S_z\!+\!1$,
$S\!\to\!S\!-\!1$). In other words, the SO coupling and the SRP
field mix the initial
$|i\rangle=\left(Q_0^{\dagger}\right)^{N}\!\!\left|0\right\rangle$ and the
final $|f_q\rangle=Q_{\bf
q}^{\dagger}\left(Q_0^{\dagger}\right)^{N\!-\!2}\!\!\!\left|0\right\rangle$
states. These ``many-exciton'' states have to be normalized (see
the normalization factors in Refs. \cite{di96,dz83}).  So,
the SRP plays the same role as the phonon field studied
previously \cite{di96}. Now the energy conservation condition
takes the form:  $2\epsilon_Z\approx {\cal E}_{q}$, where the
interaction of ``nonzero'' SE with the SRP is ignored as compared
to other members of this equation. If solving this for $q$, we
obtain $q=q_0\equiv \sqrt{2M_n\epsilon_Z}/l_B$.

The detailed calculation of the SR rate is truly similar to that
performed in Ref. \cite{di96}. Indeed, the Fourier
expansion $u({\bf r})=\sum_{{\bf q}}{\overline u}(q)e^{i{\bf
q}{\bf r}}$ looks like the phonon field created by ``frozen'' (of
zero frequency) phonons. Then this  SRP field and the SO
Hamiltonian are treated perturbatively. In so doing, it is
convenient to present them in terms of the excitonic operators. We
calculate the relevant matrix element between the $|i\rangle$
and $|f_q\rangle$ normalized states and obtain 
$\left|{\EuScript
M}_{i\to f_q}\right|^2\!\!=\!\!  N^2(\alpha^2\!+\!\beta^2)|q{\overline
u}(q)|^2/(\hbar\omega_c)^2N_{\phi}$.  Finally, again with the use
of the Fermi golden rule, we find after summation over all
states $|f_q\rangle$ that the SR rate takes the same form as in
the case of the phonon mechanism$\,$\cite{di96}, $dS_z/dt=\left(\Delta
S_z\right)^2\!/\tau_{\mbox{\tiny SRP}}N_{\phi}$, but incorporates
a different time constant:
$$
  \tau_{\mbox{\tiny SRP}}^{-1}=
  \frac{16\pi^2(\alpha^2+\beta^2)M_n^2\epsilon_Z{\overline K}(q_0)}{\hbar^3\omega_c^2l_B^4}\,.
$$
Here ${\overline K}$ stands for the Fourier component of the
correlator [the equivalence ${\overline
K}(q)\!=L^2\left|{\overline u}(q)\right|^2/4\pi^2$ has
been employed].  The SR follows the low
$$
  \Delta S_z(t)=\frac{\Delta S_z(0)}{1+t\left|\Delta S_z(0)\right|/
  \tau_{\mbox{\scriptsize sr}}N_{\phi}}\,,
$$
where $\tau_{\mbox{\scriptsize sr}}^{-1}\!=
\tau_{\mbox{\scriptsize ph}}^{-1}\!+\!\tau_{\mbox{\tiny
SRP}}^{-1}$, because the relaxations of both types proceed in
parallel. A natural question is:  {\it what is the ratio of the
times} $\tau_{\mbox{\scriptsize ph}}$ {\it and}
$\tau_{\mbox{\tiny SRP}}${\it ?} If $T\!\lesssim\!\hbar c_s q_0$ (in
fact, this means that $T\!\lesssim\!1\,$K) and $B\!<\!\!15$, then the SR
time $\tau_{\mbox{\scriptsize ph}}$ depends weakly on $T$ and
$B$.  In particular, at $\nu\!=\!1$ we find that
$\tau_{\mbox{\scriptsize ph}}\simeq 10\,\mu$s \cite{di96,di99}. The
ratio of interest is determined only by the Fourier component
${\overline K}(q_0)$:  $\tau_{\mbox{\scriptsize
ph}}/\tau_{\mbox{\tiny SRP}}\!=\!0.24\pi\tau_pp_0
\Delta^2\Lambda^2\!e^{-\Lambda^2q_0^2/4}/\hbar^2c_s$ [for
${\overline K}$ we have substituted the value calculated
with help of Eq. (1)]. So, for the actual parameters
$\tau_{\mbox{\scriptsize ph}}/\tau_{\mbox{\tiny SRP}}\sim 100-1000$,
i.e. exactly the ``disorder'' time ($\tau_{\mbox{\tiny
SRP}}\!\sim\!10^{-8}-10^{-7}\,$s) governs the breakdown of this
Goldstone mode.   In
conclusion, we remind that {\it the  relaxation is going non-exponentially and
the actual time is increased by a factor of} $N_{\phi}/\Delta
S_z(0)$.

{\bf 4.} I thank Y.B. Levinson for helpful discussion and acknowledge
support by the MINERVA Foundation and by the Russian Foundation
for Basic Research. I thank also the Weizmann Institute of
Science (Rehovot) for hospitality.

\begin{figure}[bp]
\begin{center}
\includegraphics*[width=0.8\textwidth]{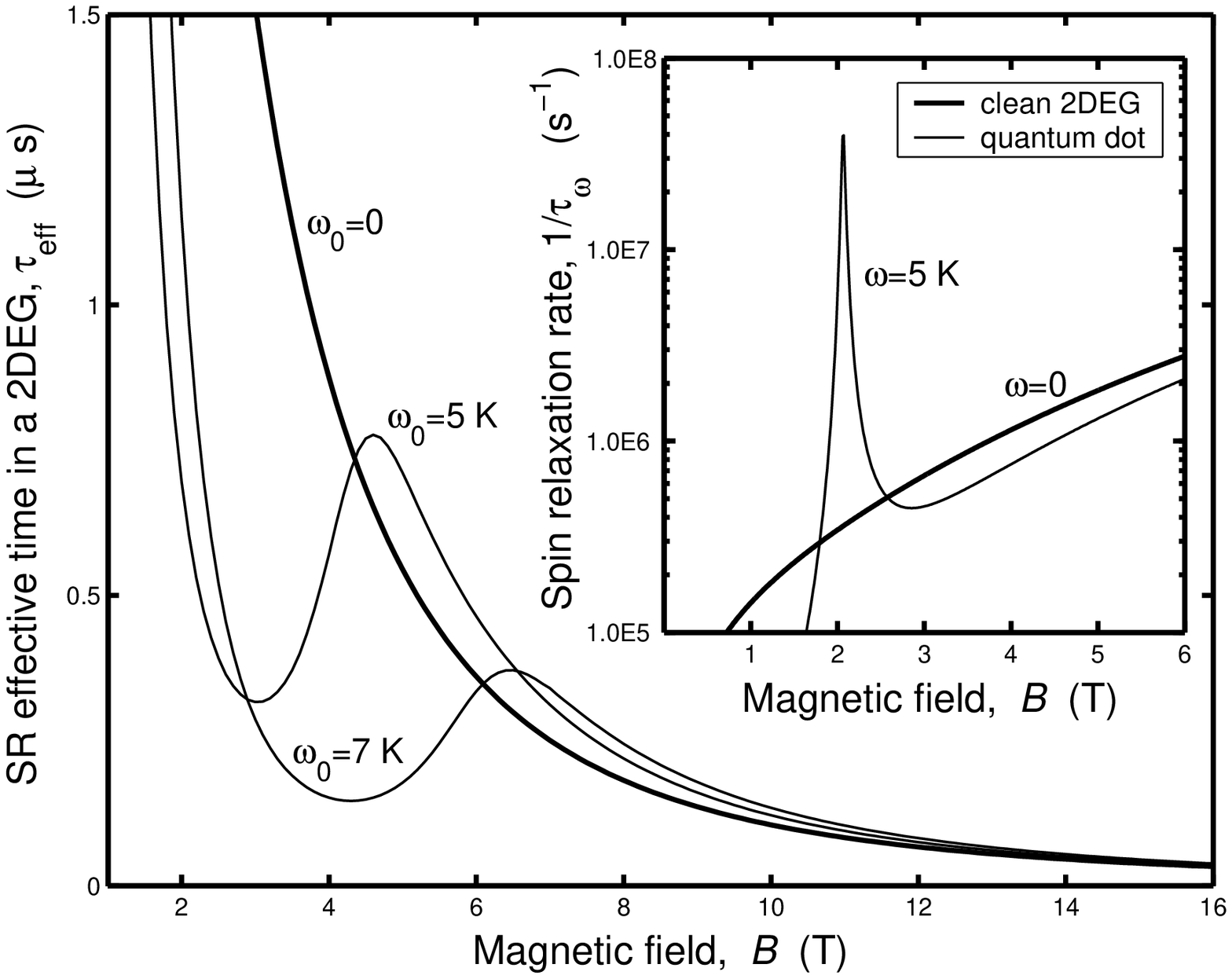}
\end{center}
\caption{Calculations are carried out for
$\alpha=\beta/3=10^{-6}\,$K$\cdot$cm
and $c_s=3.37\cdot 10^5\,$cm/s (other material parameters are given in
the text). In the inset the position of
the SR peak corresponds to the condition $\epsilon_Z=\delta$.
The evolution of $B$-dependences of the spin relaxation time
$\tau_{\mbox{\scriptsize eff}}$ with the parameter $\omega_0$ is shown
in the main picture.}
\end{figure}
\end{document}